\documentclass[twocolumn,showpacs,preprintnumbers,amsmath,amssymb]{revtex4-1}
\usepackage{graphicx}
\usepackage{amsmath}
\begin{document}

\title{Vibrational mechanics in an optical lattice: controlling transport 
via potential renormalization}

\author{A. Wickenbrock$^{1}$,  P.C. Holz$^{1}$, N.A. Abdul Wahab$^{1}$, P. Phoonthong$^{1}$, D. Cubero$^{2}$, and F. Renzoni$^{1}$}

\affiliation{$^{1}$Department of Physics and Astronomy, University College
London, Gower Street, London WC1E 6BT, United Kingdom}

\affiliation{$^{2}$
Departamento de F\'{\i}sica Aplicada I, EUP, Universidad de Sevilla,
Calle Virgen de \'Africa 7, 41011 Sevilla, Spain and
F\'{\i}sica Te\'orica, Universidad de Sevilla, Apartado de Correos
1065, Sevilla 41080, Spain}

\date{\today}

\begin{abstract}
We demonstrate theoretically and  experimentally the phenomenon of 
vibrational resonance in a periodic potential, using cold atoms in an optical lattice 
as a model system.   A high-frequency (HF)  drive, with frequency much 
larger than any characteristic frequency of the system, is applied by 
phase-modulating one of the lattice beams.  We show that the HF drive  
leads to the renormalization of the potential. We used transport measurements 
as a probe of the potential renormalization. The very same experiments also 
demonstrate that transport can be controlled by the HF drive via potential 
renormalization.
\end{abstract}


\maketitle

The control of transport is a recurrent topic in physics, chemistry and biology. 
The typical scenario corresponds to particles diffusing on a periodic substrate,
with transport controlled by the application of dc and ac external fields
\cite{reimann,rmp09}. The ultimate limit for the control of transport is often the 
impossibility of tuning  the periodic  potential, as it is usually the case in solid state.

In this work we provide a proof-of-principle of how this limitation can be overcome, 
and  demonstrate theoretically and experimentally the control of a periodic
potential amplitude 
via a strong high-frequency (HF) oscillating field. The potential is renormalized, 
with its amplitude controlled by the strength and frequency of the HF field. The 
mechanism underlying the potential renormalization is the so-called vibrational 
resonance, intially introduced \cite{landa} and observed \cite{giacomelli,casado04,casado07,cubero06} in 
bistable systems. Our experiment uses cold atoms in a dissipative optical 
lattice as a model system. However, the phenomenon demonstrated here is very 
general, and is relevant to  any classical system of particles in a 
periodic potential.  This may also offer a  possibility of tuning the 
potential in solid state systems, where this is usually considered impossible. 
Combined with previous work which showed how ac fields can be used to 
control transport via dynamical symmetry breaking \cite{reimann, rmp09} and 
tunnel coupling renormalization \cite{holthaus,arimondo}, the present work 
demonstrates that a complete control of transport can  be achieved via ac fields.

Our experimental work relies on the study of the transport properties of atoms
in an optical lattice for different strengths of the applied HF field. We will 
demonstrate that by tuning the HF field it is possible to control the amplitude of 
the potential, and to make it vanish. In this respect, the use of cold atoms in 
dissipative  optical lattices is very convenient as the transport properties in these 
systems have been studied in detail \cite{zoller,katori,robi}, and this allows us to 
use the transport measurements to characterize the potential. We will provide two 
different sets of measurements, as supporting evidence of the potential renormalization.
First, we will demonstrate that the diffusion properties, which are known to strongly 
depend on the potential depth \cite{katori,zoller}, can be controlled by the HF field 
in a way which corresponds to the potential renormalization. Second, we will show that 
also directed transport, as induced by harmonic-mixing (HM) \cite{HM} of a bi-harmonic 
drive, can be controlled by the HF field.  In fact, anharmonicity, together with the 
breaking of a dynamical symmetry, leads to the creation of directed currents in harmonic mixing. 
Therefore, whenever the potential, renormalized 
by the HF field, vanishes, directed transport should  cease \cite{fabio05,fabio06,fabio07}.

Before discussing the experimental results, we  introduce a  model useful for the
understanding of  the potential renormalization of a dissipative optical lattice as produced by 
a HF oscillating  field.  We consider the simplest model of a dissipative optical lattice: 
 a $J_g = 1/2 \to J_e = 3/2$ atom, of mass $m$,  illuminated by two counterpropagating
 laser fields with orthogonal linear polarizations. This configuration generates a 1D 
optical lattice \cite{robi}. The atom in the $\pm$ ground state experiences the potential 
$U_{\pm}(z) = U_0 [-2 \pm \cos(2kz)]/2$, where $z$ is the laser beam propagation axis, 
$k$ the laser field wavevector and $U_0$ the optical lattice depth \cite{robi,note1}. 

We now introduce a HF oscillating force with frequency $\omega_\mathrm{HF}$ and 
amplitude $A_\mathrm{HF}$: 
\begin{equation}
F_\mathrm{HF}(t) = A_\mathrm{HF} \sin(\omega_\mathrm{HF}t + \phi_0)~, \label{eq:rocking} 
\end{equation}
with $\phi_0$ a (mainly irrelevant) phase which describes the state of the oscillating force at $t=0$. 
Of interest here is the high-frequency case, where the frequency of the HF drive
is much larger then any characteristic frequency of the system, in the 
present case the vibrational frequency $\omega_v$ of the atoms at the bottom of 
the well.  In the asymptotic limit  of infinite amplitude and frequency of the drive  ($\omega_\mathrm{HF}\to \infty$,
$A_\mathrm{HF}\to \infty$), it is possible to show \cite{note2} that,  consistently with Refs. 
\cite{landa, fabio05,fabio06,fabio07}, the atomic dynamics corresponds to the 
motion in a static (i.e. without HF field) dissipative optical lattice, 
with renormalized amplitude $\tilde{U}$:
\begin{equation}
\tilde{U}_{\pm}(\hat{z}) = U_0 [-2 \pm J_0(2kr) \cos(2k\hat{z})]/2 \label{eq:renorm}
\end{equation} 
where  $J_0$ is the Bessel function of the
first kind, and $r =A_\mathrm{HF}/(m\omega^2_\mathrm{HF})$ is the parameter --here and 
thereafter termed the {\it HF ratio}-- which controls the renormalization of 
the optical lattice.

The above analysis shows that, in the asymptotic limit of infinite frequency and strength, 
a HF field leads to an effective renormalization of the potential.   We now  
consider finite values of driving away from infinity that are experimentally 
accessible. The atomic transport in the optical lattice in the presence of a HF 
field is numerically studied for  two different set-ups, which correspond to 
the ones used to provide the experimental evidence.

\begin{figure}[ht]
\begin{center}
\includegraphics[height=2.in]{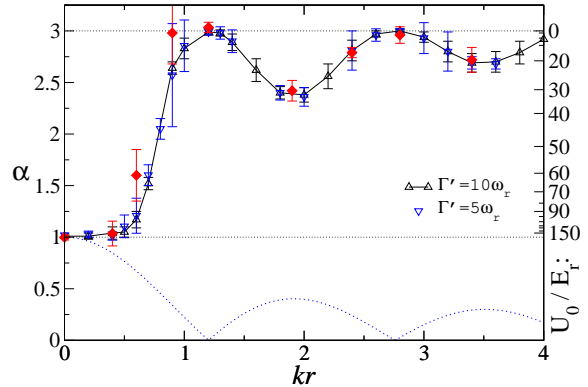}
\end{center}
\caption{
Left axis: numerical results, as obtained by Monte Carlo simulations, for the spatial diffusion 
exponent as a function of the HF ratio $r$  for an optical lattice with a depth $U_0=200 E_r$.
Right axis: value of $U_0$ which corresponds to  the exponent $\alpha$ for an 
undriven lattice.  Triangles correspond to  results  
obtained in the infinite limit,   for  a photon scattering rate  $\Gamma^\prime=5\omega_r$ and 
$\Gamma^\prime=10\omega_r$, were $\omega_r$ is the recoil 
frequency.  The filled diamonds refer to simulations with 
a HF field of finite amplitude,  with frequency 
$\omega_\mathrm{HF}/\omega_{v}=20$ and $\Gamma^\prime=10\omega_r$. 
The solid line is a guide for the eye for the results  corresponding to the infinite limit
with $\Gamma^\prime=10\omega_r$. The dotted line is 
$|J_0(2kr)|$.  The error bars on the numerical results correspond to the finite statistics of the 
Monte Carlo simulations. }
\label{fig1}
\end{figure}

In the first set-up, a HF force of finite amplitude and 
frequency  is applied to atoms in a dissipative optical 
lattice. For this scheme, the effective renormalization 
of the optical potential can be detected by studying the
diffusion properties of the atoms through the lattice. 
In fact, for a dissipative optical lattice of the type 
considered here, it is well established \cite{zoller,katori} 
that there is a critical potential depth located at about $U_{\rm cr}\sim 100 E_r$, 
with $E_r=\hbar^2k^2/(2m)$ the recoil energy, which separates 
two very different regimes. For potential depths larger than 
the critical one, the diffusion is normal. Instead, for potential 
depth lower than the critical one, the diffusion becomes anomalous, 
with the exponent of the diffusion dependent on the potential 
amplitude. To be quantitative, we define  the diffusion exponent 
$\alpha$  as $\langle x^2(t)\rangle - \langle x(t)\rangle^2 \sim t^\alpha$ 
in the limit $t \to \infty$. According to this definition, 
$\alpha = 1$ corresponds to normal diffusion, while $\alpha > 1$ 
characterizes superdiffusion. In an undriven optical lattice, 
superdiffusion is encountered at shallow optical potentials 
(below the critical depth), with the exponent $\alpha$ 
increasing for decreasing potential depth. Thus we will take the 
exponent of the diffusion as a measure of the potential 
depth.  To assess the effective renormalization of the 
potential by a HF field, we numerically simulated 
the dynamics of the atoms in a deep optical lattice with a 
HF drive. We determined the diffusion exponent as a function of 
the HF ratio $r$ which for infinite frequency and
amplitude of the drive determines the lattice renormalization.
Our results, reported in Fig.~\ref{fig1}, show that the diffusion 
exponent $\alpha$ can be controlled
by the HF field, with a dependence consistent with the potential 
renormalization derived in the infinite limit,  Eq.~(\ref{eq:renorm}): 
$\alpha$ increases whenever the potential depth is decreased, 
with the largest values of $\alpha$ produced by the $r$ values 
corresponding to the zeros of the Bessel function, i.e. to 
vanishing potentials. The upper bound of $\alpha=3$  for $U_0=0$ 
corresponds to the vanishing of the friction mechanism ({\it Sisyphus cooling} \cite{robi})
associated with the optical lattice. Figure \ref{fig1} also reports the value 
of $U_0$ which corresponds to  the exponent $\alpha$ for an 
undriven lattice, so to make explicit the correspondence between 
a driving with HF ratio $r$ and the depth of the renormalized potential.  
Finally,  we notice that our results for the diffusion exponent $\alpha$ 
essentially coincides with the values derived in the infinite limit. We
can thus conclude that the potential is effectively 
renormalised according to the dependence obtained in the 
infinite limit (see Eq.~(\ref{eq:renorm})).

%%%%%%%%%%%%%%%%%%%%%%%%%%%%%%%%%%%%%%%%%%%%%
\vspace{0.75cm}

%%%%%%%%%%%%%%%%%%%%%%%%%%%%%%%%%%%%%%%%%%%%%

\begin{figure}[ht]
\begin{center}
\includegraphics[height=2.in]{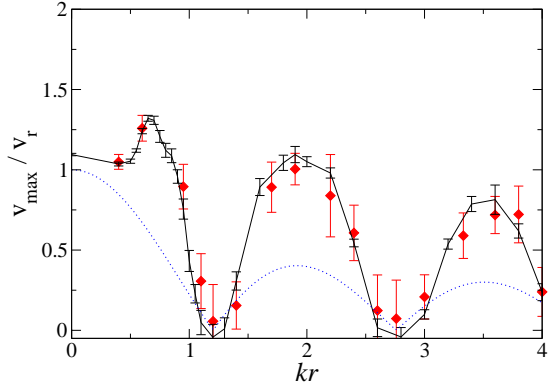}
\end{center}
\caption{Numerical results, as obtained by Monte Carlo simulations, 
for the amplitude of the current $v_{\rm max}$, rescaled by the recoil velocity 
$v_r$, as a function of the HF ratio $r$ under a biharmonic driving force of the 
form of  Eq.~\protect(\ref{eq:biharmonic}) with $A_1=A_2=1$,  
$F_0=140 \hbar k \omega_r$ and $\omega=\omega_v$. The solid line corresponds 
to the results obtained in the infinite limit   and the diamonds to the simulation 
results  with  $\omega_{\rm HF}/\omega_v=20$,  both cases with 
$\Gamma^\prime=10\omega_r$. The dotted line is $|J_0 (2kr)|$. The error bars on 
the numerical results correspond to the finite statistics of the Monte Carlo simulations. 
}
\label{fig2}
\end{figure}

In the second set-up, besides the HF drive, a bi-harmonic  force of the form 
\begin{equation}
F(t) = F_0 [ A_1\cos(\omega t) +  A_2\cos(2\omega t+\phi) ]
\label{eq:biharmonic}
\end{equation}
is also applied to the atoms in the lattice. Here $\omega$ 
is the frequency of the drive, of the same order of magnitude 
or smaller than the vibrational frequency, and $\phi$ the 
relative phase between harmonics. The amplitude of the 
lattice, and its renormalization by the HF field, can be 
determined by observing the directed motion of the atoms 
through the lattice. In fact, the two harmonics of the 
drive are mixed by the nonharmonic potential, thus producing 
directed motion of the atoms through the lattice \cite{advances}.
The average current being proportional to the nonharmonicity of the potential, directed transport
measurements give access to the potential amplitude. More precisely, for weak driving the average atomic 
velocity is expected to be of the form $v=v_{\rm max}\sin(\phi-\phi_d)$ , with $\phi_d$ a dissipation-induced phase lag  \cite{advances}. In our simulations we 
determined the velocity $v$ for different values of the phase $\phi$, so to derive the maximum
velocity $v_{\max}$. Then by varying the strength of the HF drive, we were able to determine 
$v_{\max}$ as a function of the $r$-parameter, with results as in Fig.~\ref{fig2}.
Once again, the results produced with a field of large,
but finite, frequency and amplitude essentially coincide
with those obtained in the infinite limit. These results 
also show that current measurements can be used to probe
the potential renormalization. Whenever the HF field leads
to a shallower potential, as from Eq.~(\ref{eq:renorm}), the
current is reduced, with zero current observed for those 
values of $r$ leading to a vanishing potential.

%%%%%%%%%%%%%%%%%%%%%%%%%%%%%%%%%%%%%%%%%%%%%
\vspace{0.75cm}

%%%%%%%%%%%%%%%%%%%%%%%%%%%%%%%%%%%%%%%%%%%%%

Our experimental demonstration of potential renormalization 
via HF field relies on the two detection schemes outlined 
above. In both set-ups, $^{87}$Rb atoms are cooled and 
trapped in a magneto-optical trap (MOT). After a compression 
phase of 50 ms, and 8 ms of optical molasses, the atoms are 
loaded into a 1D dissipative optical lattice.
The lattice is created by the interference of two 
linearly polarized and counter-propagating laser beams, 
red detuned from resonance with the
D$_2$-line $F_g=2\to F_g=3$ atomic transition. 
One of the lattice beams is sent through a double pass 
electro-optical modulator (EOM), so to be able to apply a 
HF phase-modulation. In the reference frame of the lattice, 
such a phase modulation translates into a rocking force of the 
form of  Eq.~(\ref{eq:rocking}).  Quantitatively, a phase 
modulation $\alpha (t)$ leads to a force $F(t)=m\ddot{\alpha}(t)/(2k)$
in the reference frame of the lattice \cite{advances}. 
In the experiments,  the modulation is progressively 
turned on starting after 1 ms equilibration time in the optical lattice, 
with a turn-on ramp of  1 ms. Thereafter the procedure differed 
for the two set-ups.

In the first experiment, we study the diffusion of the atoms in the 
optical lattice in the presence of the HF drive. The width of the atomic 
cloud is measured by fluorescence imaging after diffusive expansion 
inside the driven lattice. The width is  measured at a fixed set of
expansion times within the interval between 1 ms and 16 ms from the 
lattice turn-on. The time range over which images are taken is limited 
by the atom loss, particularly important for the values of the HF ratio 
$r$ leading to a vanishing renormalized potential. Measurements 
of the spatial width of the atomic cloud on such a short temporal 
range do not allow us to derive an accurate value of the exponent 
of the diffusion $\alpha$ \cite{note3}. Instead, we characterize the diffusion by an 
effective diffusion coefficient $D$, as obtained by fitting the data with 
$\langle x^2(t)\rangle - \langle x(t)\rangle^2 = 2 D t $. 
Clearly, superdiffusion leads to a large enhancement
of the derived effective diffusion coefficient. Thus 
a large increase in the effective diffusion coefficient 
can be taken as signature of the reduction of the potential, 
as produced by the renormalization by the HF field.

%%%%%%%%%%%%%%%%%%%%%%%%%%%%%%%%%%%%%%%%%%%%%
\vspace{0.75cm}
\begin{figure}[ht]
\begin{center}
\includegraphics[height=2.5in]{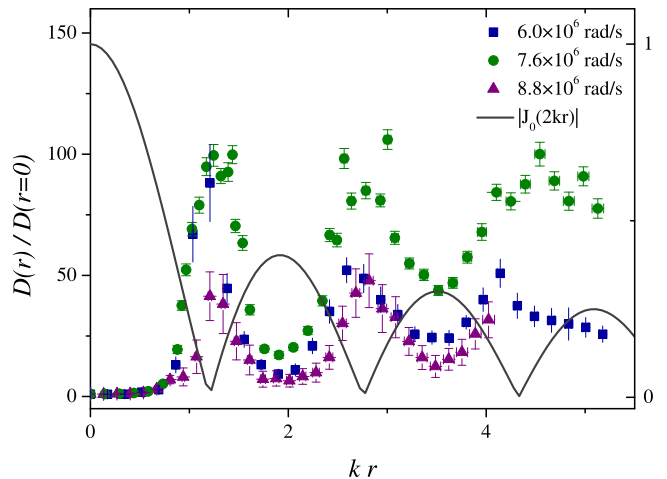}
\end{center}
\caption{Experimental results for the effective diffusion 
coefficient $D$ as a function of the HF ratio $r$ for 
different values of $\omega_{\rm HF}$, as indicated in 
the figure. The data are rescaled by the value of the 
diffusion constant for an undriven lattice. The vibrational 
frequency of the atoms at the bottom of the well,  
as determined by measuring the lattice beam power and waist,   is
$\omega_v=(9\pm 1)\cdot 10^5$ rad/s.
The solid line,  with values on the right axis, is $|J_0(2kr)|$.
}
\label{fig3}
\end{figure}
%%%%%%%%%%%%%%%%%%%%%%%%%%%%%%%%%%%%%%%%%%%%%

Our experimental results for the effective diffusion coefficient as a function of the HF ratio $r$ are
reported in Fig.~\ref{fig3}. The data clearly show that the atomic diffusion is significantly modified
by the HF drive, with a dependence of the effective diffusion coefficient on the HF ratio $r$ consistent
with the potential renormalization (see e.g. Eq.~(\ref{eq:renorm}) for the analytic expression in the
limit of infinite frequency and amplitude). Indeed, the effective diffusion coefficient increases
whenever the HF ratio $r$ corresponds to decreasing depth of the optical lattice, with the largest
values of the diffusion constant observed in correspondence of the values of $r$ leading to
a vanishing (in the infinite limit) optical lattice. This shows that the HF drive renormalizes the
optical potential, in agreement with the general theory \cite{landa} and with our numerical analysis for
the specific system.

\begin{figure}[ht]
\begin{center}
\includegraphics[height=2.5in]{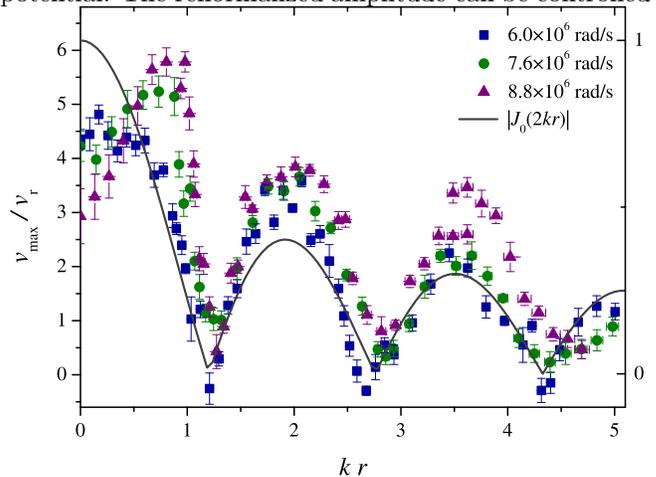}
\end{center}
\caption{Experimental results for the amplitude of the current $v_{\rm max}$,
rescaled by the recoil velocity $v_r$, of directed transport through 
the optical lattice as a function of the HF ratio $r$. 
In addition to the HF drive, a biharmonic force of the form of 
Eq.~\protect(\ref{eq:biharmonic}) is applied to the atoms, with 
parameters $A_1=1$, $A_2=2$, $\omega=9.42\cdot 10^4$ rad/s, 
$F_0=112 \hbar k \omega_r$.  The vibrational frequency of the atoms at 
the bottom of the well is $\omega_v=(9\pm 1)\cdot 10^5$ rad/s.
The solid line, with values on the right axis, is $|J_0(2kr)|$.}
\label{fig4}
\end{figure}

In the second experiment, we probe the amplitude of the renormalized potential by studying 
directed transport following harmonic mixing of two harmonics, as outlined in the numerical analysis. 
With respect  to the previous experiment devoted to the study of the atomic diffusion, an additional  
bi-harmonic drive, with frequencies $\omega$, $2\omega$ and phase difference $\phi$ is introduced. 
This is done using additional acousto-optical modulators (AOMs). In the reference frame of the lattice, 
the bi-harmonic phase modulation corresponds to a driving force of the form of Eq.~(\ref{eq:biharmonic}).  
In the experiment, the  HF driving is first ramped up, as in the previous experiment. Then  
the biharmonic drive is progressively turned on with a ramp-up time of 4  ms. The velocity of 
the center-of-mass of the atomic cloud is derived by position measurements obtained via
fluorescence imaging. The measurements are repeated for 10 different values of the phase 
difference $\phi$ between harmonics.  The data are then fitted by the expected dependence 
$v=v_{\rm max}\sin(\phi-\phi_d)$,  thus deriving a value for $v_{\rm max}$ which can be taken as  
a measure of the renormalized potential depth. In fact, the mixing of harmonics requires an
anharmonic potential, with the current generated proportional to the anharmonicity.  Our results for
$v_{\rm max}$ as a function of the HF ratio $r$ are presented in Fig.~\ref{fig4}. These data for the 
directed transport amplitude are consistent with the renormalization of the potential depth 
by the HF drive. In fact, whenever the value of the HF ratio $r$ corresponds to a reduced potential
depth, the current decreases, with zero current observed for the $r$-values corresponding to the 
zeros of the Bessel function, a signature of the vanishing optical lattice. These results also demonstrate
a new scheme for the control of the transport via ac fields: the amplitude of the current can be controlled
by a variation in  the HF field and the direction reversed via a $\pi$-shift in the relative phase between harmonics. 
Finally,  we notice that there is
 a small deviation, both in the experiment and in the numerical simulations (see Fig.~\ref{fig2}) 
from the behaviour expected from the Bessel function  at small values of $r$,  with the data
showing an extra peak at $kr\sim 0.75$. This peak could be explained by a superimposed resonance 
corresponding to the matching of the frequency of the biharmonic force with the oscillation frequency 
of the atoms at the bottom of the  renormalized well.  

%%%%%%%%%%%%%%%%%%%%%%%%%%%%%%%%%%%%%%%%%%%%%
\vspace{0.75cm}

%%%%%%%%%%%%%%%%%%%%%%%%%%%%%%%%%%%%%%%%%%%%%

In conclusion, in this work we demonstrated experimentally the phenomenon of vibrational resonance
in a dissipative optical lattice. The application of a HF drive, with frequency much larger than any
characteristic frequency of the system, leads to the renormalization of the potential. The renormalized
amplitude can be controlled by the HF drive parameters. We used transport measurements as a probe
of the potential renormalization. The very same experiments also demonstrated that transport 
can be controlled by the HF drive via potential renormalization.

The possibility to renormalize a potential via ac fields,  as demonstrated here,  is very general, and it
is applicable to any system of particles in a periodic potential. As such, it paves the way to the control 
of potentials in systems in which they are not directly accessible,  and it may also be applicable to 
solid state systems where ac drives can be introduced  by the application of electric fields. 

Finally, our set-up can also be taken as the demonstration of a sensor able to detect signals with frequency 
exceeding any internal frequency of the sensor \cite{fabio05,fabio06,fabio07}. Here, the signal 
detected is the HF drive whose presence, although not coupling to any internal mode of the system,
can be precisely detected due to its effect via the potential renormalization.

We acknowledge financial support from the Leverhulme Trust,  the Ministerio
de Ciencia e Innovaci\'on of Spain FIS2008-02873 (DC), and the DAAD (P.C. H).


\begin{thebibliography}{99}
\bibitem{note3} Atom losses lead to an underestimate of the diffusion exponent, as many super-diffusing
atoms are lost.
\end{thebibliography}
\end{document}

\title{Vibrational mechanics in an optical lattice: controlling transport 
via potential renormalization -- Supplemental Material}

\author{A. Wickenbrock,  P.C. Holz, N.A. Abdul Wahab, P. Phoonthong, D. Cubero, and F. Renzoni}

\maketitle

%%%%%%%%%%%%%%%%%%%%%%%%%%%%%%%%%%%%%%%%%%%%%%%%%%%%%%%%%%%%%%%%%%%%%%%%
\section{Model and definitions}
%%%%%%%%%%%%%%%%%%%%%%%%%%%%%%%%%%%%%%%%%%%%%%%%%%%%%%%%%%%%%%%%%%%%%%%%
As a model for our experiment, we consider the simplest configuration in which Sisyphus cooling has been shown to take place,
 i.e. the case of a $J_g=1/2\rightarrow J_e=3/2$ atomic transition in a 1D optical lattice generated by two counterpropagating 
laser fields with orthogonal linear polarizations. This is the so-called lin~$\perp$~lin configuration \cite{robi}. In Ref.~\cite{petsas}, 
the following generalized Fokker-Planck equation was found in the semiclassical limit for the probability density $P_{\pm}(z,p,t)$ of
each atom that is in the ground state sublevel $|\pm\rangle=|J_g=1/2,M_g=\pm1/2\rangle$ at the position $z$ with momentum $p$:
\begin{eqnarray}
\left[ \frac{\partial}{\partial t} + \frac{p}{m}\frac{\partial}{\partial z} 
-U_{\pm}^\prime(z)\frac{\partial}{\partial p}+F(t)\frac{\partial}{\partial p}\right] P_{\pm}= \nonumber\\
-\gamma_{\pm}(z)P_{\pm}+\gamma_{\mp}(z)P_{\mp} \nonumber\\
+\frac{\partial^2}{\partial p^2}\left[D_{\pm}(z)P_{\pm}+L_{\pm}(z)P_{\mp}\right],
\label{eq:fp}
\end{eqnarray}
where $m$ is the atomic mass and $U_{\pm}^\prime(z)=dU_{\pm}(z)/dz$;
\begin{equation}
U_{\pm}(z)=\frac{U_0}{2}[-2\pm\cos(2kz)]
\end{equation}
is the optical bipotential created by the laser fields, with $k$  the laser field wave vector; $F(t)$ is a 
time-dependent driving force that can be generated by phase modulating one the lattice beams \cite{advances};
\begin{equation}
\gamma_{\pm}(z)=\frac{\Gamma^\prime}{9}[1\pm\cos(2kz)]
\end{equation}
is the transition rate between the ground state sublevels, with $\Gamma^\prime$  the  photon scattering rate;
\begin{equation}
D_{\pm}(z)=\frac{7\hbar^2k^2\Gamma^\prime}{90}[5\pm\cos(2kz)]
\end{equation}
is a noise strength coefficient describing the random momentum jumps 
that result from the interaction with the photons without transition between ground state sublevels;
 and
\begin{equation}
L_{\pm}(z)=\frac{\hbar^2k^2\Gamma^\prime}{90}[6\mp\cos(2kz)]
\end{equation}
is related to random momentum jumps that appear in fluoresecence cycles when the atom undergoes a transition between the atomic sublevels. The normalization condition is given by
\begin{equation}
\int \mathrm{d}z\int\mathrm{d}p \, [P_-(z,p,t)+P_+(z,p,t)]=1.
\end{equation}
   
We consider two different types of time-dependent driving forces:  a bi-harmonic drive of the form
\begin{equation}
F_d(t)=A_1\cos(\omega t)+A_2\cos(2\omega t+\phi),
\label{eq:bihdriving}
\end{equation}
and a high-frequency (HF) drive of the form
\begin{equation}
F_{\mathrm{HF}}(t)=A_{\mathrm{HF}}\sin(\omega_{\mathrm{HF}}t+\phi_0).
\label{eq:HF}
\end{equation} 

The bi-harmonic drive is used to probe the potential amplitude.
The HF drive determines the potential renormalization, as discussed in next
Section.

%%%%%%%%%%%%%%%%%%%%%%%%%%%%%%%%%%%%%%%%%%%%%%%%%%%%%%%%%%%%%%%%%%%%%%%%
\section{Potential renormalization by high-frequency driving}
%%%%%%%%%%%%%%%%%%%%%%%%%%%%%%%%%%%%%%%%%%%%%%%%%%%%%%%%%%%%%%%%%%%%%%%%
We now study the effect  on the cold atom system of a high-frequency signal $F_{\mathrm{HF}}$, 
of the form of Eq. \ref{eq:HF} with $\phi_0$ an arbitrary initial phase.
A low-frequency bi-harmonic drive $F_d(t)$ is also included in the analysis, to model 
the experiments in which the potential is probed by using this type of driving.

We are interested in situations in which the frequency $\omega_{\mathrm{HF}}$ is
much larger than $\omega$ and any other characteristic frequency in the system. 
Formally, this can be achieved by taking the asymptotic limit $\omega_{\mathrm{HF}}\rightarrow\infty$.
 In this limit, it is also necessary that $A_{\mathrm{HF}}\rightarrow \infty$ if the HF signal
is to have any effect. Due to this strong driving, the momentum changes very rapidly, 
since its time-derivative is of order $A_{\mathrm{HF}}$. Integrating this dominant term in 
time, we find a rapidly changing contribution to the position $z(t)$ that goes as $-r\sin(\omega_{\mathrm{HF}}t+\phi_0)$, where 
\begin{equation}
r=\frac{A_{\mathrm{HF}}}{m\omega_{\mathrm{HF}}^2}.
\end{equation}
Formally, we will consider the asymptotic limit $\omega_{\mathrm{HF}}, A_{\mathrm{HF}}\rightarrow\infty$ while keeping $r$ fixed.
By extracting the fast dependence from $z(t)$,
\begin{equation}
\hat{z}(t)=z(t)+r\sin(\omega_{\mathrm{HF}}t+\phi_0),
\label{eq:zhat}
\end{equation}
it is expected that $\hat{z}(t)$ changes on a much slower time-scale than that of the HF signal. The density probabilities for the new variable are then given by 
\begin{eqnarray}
\hat{P}_\pm(\hat{z},\hat{p},t)&=&P_\pm[\hat{z}-r\sin(\omega_{\mathrm{HF}}t+\phi_0),
{}\nonumber\\
&& {}\hat{p}-rm\omega_{\mathrm{HF}}\cos(\omega_{\mathrm{HF}}t+\phi_0),t],
\end{eqnarray}
where $\hat{p}=m\,d\hat{z}/dt$.
The corresponding generalized Fokker-Planck equations are 
\begin{eqnarray}
\left[ \frac{\partial}{\partial t} + \frac{\hat{p}}{m}\frac{\partial}{\partial \hat{z}} 
-\hat{U}_{\pm}^\prime(\hat{z},t)\frac{\partial}{\partial \hat{p}}+F_d(t)\frac{\partial}{\partial \hat{p}}\right] \hat{P}_{\pm}= \nonumber\\
-\hat{\gamma}_{\pm}(\hat{z},t)\hat{P}_{\pm}+\hat{\gamma}_{\mp}(\hat{z},t)\hat{P}_{\mp} \nonumber\\
+\frac{\partial^2}{\partial \hat{p}^2}\left[\hat{D}_{\pm}(\hat{z},t)\hat{P}_{\pm}+\hat{L}_{\pm}(\hat{z},t)\hat{P}_{\mp}\right],
\label{eq:fk:2}
\end{eqnarray}
where $\hat{U}_{\pm}^\prime(\hat{z},t)=U_{\pm}^\prime[\hat{z}-r\sin(\omega_{\mathrm{HF}}t+\phi_0)]$, $\hat{\gamma}_{\pm}(\hat{z},t)=\gamma_{\pm}[\hat{z}-r\sin(\omega_{\mathrm{HF}}t+\phi_0)]$, and similarly for $\hat{D}_\pm(\hat{z},t)$ and $\hat{L}_\pm(\hat{z},t)$. These coefficients depend on time only through the HF signal. On the other hand, both $\hat{P}_\pm$ and $F_d$ vary with time on a much longer timescale. Therefore, we could remove the time dependence from the above coefficients by integrating over a time interval that includes many HF periods but in which  $\hat{P}_\pm$ and $F_d$ do not appreciably change. Equivalently, we can eliminate that fast dependence by noting that $\hat{P}_\pm$ should be independent of the HF phase $\phi_0$. Integrating Eq.~(\ref{eq:fk:2}) over the phase $\phi_0$, we finally find a generalized Fokker-Planck equation analogous to (\ref{eq:fk:2}) but with the following coefficients:
\begin{eqnarray}
\bar{U}_\pm(\hat{z})&=&\frac{1}{2\pi}\int_0^{2\pi}\!\!\!\mathrm{d}\phi_0\,U_{\pm}[\hat{z}-r\sin(\omega_{\mathrm{HF}}t+\phi_0)]\nonumber\\
&=&\frac{U_0}{2}[-2\pm\mathrm{J}_0(2kr)\cos(2k\hat{z})], \label{eq:meancoef:1}\\
\bar{\gamma}_{\pm}(\hat{z})&=&\frac{\Gamma^\prime}{9}[1\pm\mathrm{J}_0(2kr)\cos(2k\hat{z})],\label{eq:meancoef:2}\\
\bar{D}_{\pm}(\hat{z})&=&\frac{7\hbar^2k^2\Gamma^\prime}{90}[5\pm\mathrm{J}_0(2kr)\cos(2k\hat{z})],\label{eq:meancoef:3}\\
\bar{L}_{\pm}(\hat{z})&=&\frac{\hbar^2k^2\Gamma^\prime}{90}[6\mp\mathrm{J}_0(2kr)\cos(2k\hat{z})],\label{eq:meancoef:4}
\end{eqnarray}
where 
\begin{equation}
\mathrm{J}_0(2kr)=\frac{1}{2\pi}\int_0^{2\pi}\!\!\!\mathrm{d}\phi_0\,\cos(2kr\sin\phi_0)
\label{eq:bessel}
\end{equation}
is the Bessel function of the first kind. Eqs.~(\ref{eq:meancoef:1})--(\ref{eq:meancoef:4})
describe the system renormalization by the HF field in the asymptotic limit $\omega_\mathrm{HF}\to \infty$. Equivalently, it can also be seen as the lowest order of a multiple time-scale formalism using the expansion parameter $\varepsilon=\omega/\omega_\mathrm{HF}$ (see for example \cite{casado10}).